# Simultaneous Acquisition of Multi-nuclei Enhanced NMR/MRI by Solution State Dynamic Nuclear Polarization

Yugui He,[1,2] Zhen Zhang,[1,3] Jiwen Feng,[1] Chongyang Huang,[1] Fang Chen,[1] Maili Liu,[1] Chaoyang Liu[1,2]

[1] State Key Laboratory of Magnet Resonance and Atomic and Molecular Physics (Wuhan Institute of Physics and Mathematics, Chinese Academy of Sciences), Wuhan 430071, China

[2] School of Optical and Electronic Information, Huazhong University of Science and Technology, Wuhan 430074, China

[3] University of Chinese Academy of Sciences, Beijing 100048, China

E-mail: jwfeng@wipm.ac.cn, chyliu@wipm.ac.cn

**ABSTRACT:** Dynamic nuclear polarization (DNP) has become a very important hyperpolarization method because it can dramatically increase the sensitivity of nuclear magnetic resonance (NMR) of various molecules. Liquid-state DNP based on Overhauser effect is capable of directly enhancing polarizations of all kinds of nuclei in the system. The combination of simultaneous Overhauser multi-nuclei enhancements with the multi-nuclei parallel acquisitions provides a variety of important applications in both MR spectroscopy (MRS) and image (MRI). Here we present two simple illustrative examples for simultaneously enhanced multi-nuclear spectra and images to demonstrate the principle and superiority. We have observed very large simultaneous DNP enhancements for different nuclei, such as $^1$H and $^{23}$Na, $^1$H and $^{31}$P, $^{19}$F and $^{31}$P, especially for the first time to report sodium ion enhancement in liquid. We have also obtained the simultaneous imaging of $^{19}$H and $^{31}$P at low field by solution-state DNP for the first time. This method can obtain considerably complementary structure-determination information of miscellaneous biomolecules from a single measurement. It can also be used in combination with the fast acquisition schemes and quantitative analysis with reduced scan time.

## 1. Introduction

Increasing the sensitivity of nuclear magnetic resonance (NMR) experiments is always an ongoing research direction in the NMR area which can significantly broaden NMR applications, especially for molecular and cellular imaging. Hyperpolarization methods including chemically induced dynamic nuclear polarization (CIDNP) [1-2], para-hydrogen-induced polarization (PHIP) [3-4], spin-exchange optical pumping (SEOP) [5-6], dynamic nuclear polarization (DNP) [7-8] have opened a new avenue in the field of NMR. They can enhance the sensitivity by several orders of magnitude where the nuclear polarization is greatly increased by the manipulation of spin states. In those methods, CIDNP and PHIP are generated by spin-sensitive chemical reactions and it has so far remained limited to specialized chemical systems and nuclei. SEOP are achieved by optical pumping with circularly polarized laser light which is a way of hyper-polarizing noble gases such as helium, neon, krypton, and xenon only. These three methods all have some drawbacks either complex experimental conditions (low temperature, laser polarization), or only for specific circumstances (a specific chemical reaction or nucleus). DNP is an electron-nuclear double resonance technique, where the spin polarization of high-$\gamma$ electrons is transferred to the surrounding nuclei to enhance the NMR sensitivity using microwave (MW) irradiation under general ambient conditions. In general, DNP can be utilized to enhance the polarizations of all kinds of nuclei, such as $^1$H, $^{19}$F, $^{13}$C, $^{31}$P, $^{23}$Na.

Magnetic resonance imaging (MRI), as a non-radioactive and non-invasive means, has become an important tool in clinical diagnostics and biomedical research. MRI, in most case, is based on the proton signal, because the other nuclei signals of biological small molecules, for example, $^{13}$C and $^{15}$N are too weak to be imaged. DNP enhancement is an important method to enhance the biological small molecules MRI sensitivity *in vitro* and *in vivo* [9-11]. DNP-MRI molecular imaging technologies can provide the unique information to study the generation and development mechanism of the disease. Especially, it can diagnose and treat of disease by imaging special molecules. For example, we can obtain the spatial distribution and the changes over time of several related metabolic disease molecules by direct imaging. Such an important information is difficult to yield by other microscopic molecular imaging methods.

Parallel acquisition with multiple receivers was originally introduced into MRI to reduce the experiment time [12,13]. The multiple receiver coils which cover the object surface in MRI systems usually detect the same nuclear species at different spatial regions. Parallel acquisition of NMR signals for two, three, or multiple different nuclear species has proven its enormous potential for unambiguous structure determination of organic molecules and fast multidimensional NMR spectra [14-16]. A new type of experiment, named PANACEA (parallel acquisition NMR and all-in-one combination of experimental applications) allows structure determination from a single measurement and includes an internal field/frequency correction routine [17,18].

DNP mechanisms in the dielectric solid state and liquid state are different. For dielectric solid where the paramagnetic centers (free radicals or unpaired electrons) are fixed, there are three main enhancement processes: solid effect (SE) [19], cross effect (CE) [20] and thermal mixing (TM) [21]. On the other hand, DNP in liquid and metal where the paramagnetic centers (radicals) are movable is achieved through Overhauser effect (OE) [7,22]. To understand different DNP enhancement mechanisms in solid and liquid, we assume here a 2 spin coupled system of electron and nuclear spins resulting in the standard four-energy level model (Fig. 1a). The maximum DNP enhancement of the NMR signal for the nuclei with frequency $\omega_n$ in solid is obtained by irradiating double or zero quantum transition with microwave frequencies at $\omega_e+\omega_n$ or $\omega_e-\omega_n$. Clearly, solid DNP enhancement is highly selective of nuclear resonance frequency, and the microwave irradiating with frequencies $\omega_e+\omega_n$ can only enhance the polarization of one kind of nuclei with frequency $\omega_n$ when the EPR line width is narrow. However, the enhancement of Overhauser DNP effect in liquid is yielded by irradiating electron single-quantum transition with electron Larmor frequency $\omega_e$. The maximum Overhauser DNP enhancement as shown in Fig. 1b appears at the same MW frequency $\omega_e$ for all different kinds of nuclei and is thus independent of nuclear frequency. That is to say, the polarizations of all different kinds of nuclei can be enhanced simultaneously by saturating electron transition.

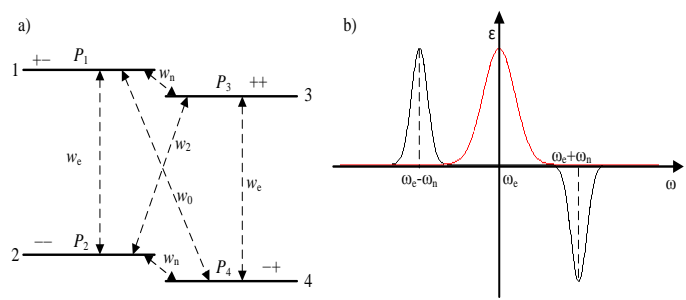

**Fig. 1.** Four level scheme of a coupled electron (S = -1/2)-nucleus (I = 1/2) spin pair (a). Nuclear polarization enhancement as a function of the microwave frequency (b), the maximum DNP in solid is at $\omega_e+\omega_n$ and $\omega_e-\omega_n$ (black), in liquid is at $\omega_e$ (red).

This fascinating advantage of simultaneous multiple nuclear Overhauser enhancements, however, has not been exploited experimentally up to now. The simultaneous multiple nuclear DNP enhancements once combined with parallel acquisition method will extremely promotes NMR developments in both MR spectroscopy (MRS) and MR image (MRI). In this paper, by parallel acquisition we have observed simultaneously DNP-enhanced spectra of different nuclei, such as $^1$H and $^{23}$Na, $^1$H and $^{31}$P, $^1$H and $^{13}$C, $^{19}$F and $^{31}$P, especially for the first time to report sodium ion enhancement in liquid. The first phosphorous image at low field by solution state DNP, together with parallel $^1$H image, is also obtained, which may advance the development of MRI of X nuclei (non-proton) at low field.

## 2. Materials and methods

To demonstrate the principle and superiority of simultaneous acquisition of Overhauser-enhanced multi-nuclei spectra and images in solution, here we present two simply illustrative examples of two-nuclei co-acquisition on our homebuilt multi-channel DNP spectrometer at 0.35T [23], although the method is clearly applicable to more sophisticated experiments involving three or more different nuclear species. After the saturation pulse of MW at electron Larmor frequency, all of the conventional pulse sequences can be used in each channel. Signals from different channels can be separately acquired either simultaneously or at different stages of their respective pulse sequences. A series of in-house-built double resonance probes are used for various nuclei with simultaneous acquisition in NMR and MRI experiments. Fig. 2 shows the simplest pulse sequences for NMR spectra (a) and images (b) used in this work.

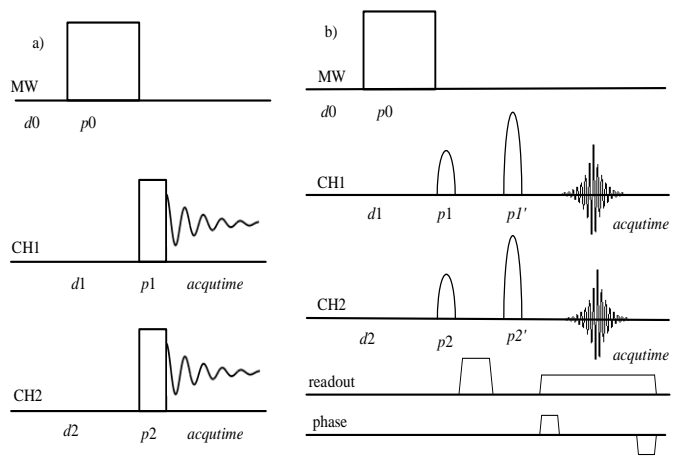

**Fig.2.** Two-channel acquired pulse sequences used in this work. (a) π/2 pulse enhancement sequence; (b) 2D spin echo image enhancement sequence; p1, p2: π/2 pulse, p1', p2': π pulse.

Here, we present the experimental data obtained simultaneously from the samples in Table 1.

Table 1 samples used in this work.

|  | Sample 1 ($^1$H and $^{31}$P) | Sample 2 ($^1$H and $^{31}$P) | Sample 3 ($^1$H and $^{23}$Na) | Sample 4 ($^{19}$F and $^{31}$P) |
|---|---|---|---|---|
| radicals | BDPA (60mM) | BDPA (90mM) | TEMPOL (120mM) | BDPA (62mM) |
| solute | 30 μL PCl$_3$ | 50 mg (C$_6$H$_5$)$_3$P | NaCl | 40μL C$_6$F$_6$, 80 mg (C$_6$H$_5$)$_3$P |
| solvent | 20 μL C$_6$H$_6$ | 60 μL C$_6$H$_6$ | H$_2$O | 50 μL C$_6$D$_6$ |
| sample volumes | 30μL | 30μL | 25μL | 30μL |

The detail processes for samples configuration are described in following:

Sample 1 for $^1$H and $^{31}$P resonances: the free radicals α, γ-bisdiphenylene-β-phenylallyl (BDPA complex with benzene (1:1)) was dissolved directly into 20 μL benzene. And then 30 μL PCl$_3$ is dissolved into the mixed solution. The concentration of free radicals is about 60mM.

Sample 2 for $^1$H and $^{31}$P resonances: The same free radicals BDPA was dissolved directly into 60 μL benzene. And then 50 mg triphenyl phosphorous ((C$_6$H$_5$)$_3$P, PPh$_3$) is dissolved into the mixed solution. The concentration of free radicals is about 90mM. Volumes of approximately 30μL of sample 1 and sample 2 were loaded into each 1.7 mm inner diameter glass capillaries and sealed and used for both DNP spectra and image measurements.

Sample 3 for $^1$H and $^{23}$Na resonances: the concentrations of NaCl is 4.8M and the free radical is TEMPOL with 120mM, volumes of approximately 25μL were loaded into 1 mm inner diameter glass capillaries and sealed.

Sample 4 for $^{19}$F and $^{31}$P resonances: The free radicals α,γ-bisdiphenylene-β-phenylallyl (BDPA complex with benzene(1:1)) was dissolved directly into 50 μL deuterated benzene and 40μL hexafluorobenzene. And then 80 mg triphenyl phosphorous (($C_6H_5$)$_3$P, PPh$_3$) is dissolved into the mixed solution. The concentration of free radicals is about 62 mM. Volumes of approximately 30μL were loaded into 1.7 mm inner diameter glass capillaries and sealed.

Hereafter in the text, the samples will be referred to as sample 1, sample 2 and sample 3, sample 4 without mentioning the concentrations of free radicals and the kinds of the solute. The structures of several molecules used in our studies for $^1$H and $^{31}$P are shown in Fig.3. This structures is determined the J-coupling effect between $^1$H and $^{31}$P in PPh$_3$ (Sample 2) but not in PCl$_3$ (Sample 1).

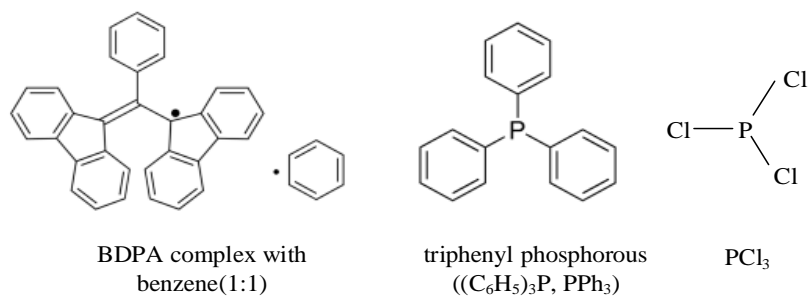

**Fig.3.** Structure of various molecules used in our studies for $^1$H and $^{31}$P

## 3. Result

### 3.1 $^1$H and $^{31}$P spectra

Phosphorus is especially interesting because of its chemical diversity of different oxidation states in organic solvent molecules and its significance in biological materials. The DNP $^{31}$P enhancement of a number of phosphorus compounds has been reported early by Potenza et al [24]. Fig.4 is the simultaneously acquired $^{31}$P and $^1$H spectra of sample 1 with (Fig.4a and b) and without (Fig.4c and d) DNP enhancement at one scan using the pulse sequence in FIG.2a. The time of microwave saturation is 2s and microwave power is 8W (MW resonator input terminal). We observe a large $^1$H signal enhancement of up to 60 fold in benzene which is determined from the ratio of the areas of the spectra. We also observe an intensive Overhauser-enhanced $^{31}$P signal at MW irradiation but such a $^{31}$P signal is not observable without MW irradiation even it was accumulated for 1024 scans. The normalized $^{31}$P DNP enhancement factor ε of sample 1 as a function of the microwave power is shown in Fig.5. It is indicated that the electron transitions have been completely saturated when the microwave power is about 8W.

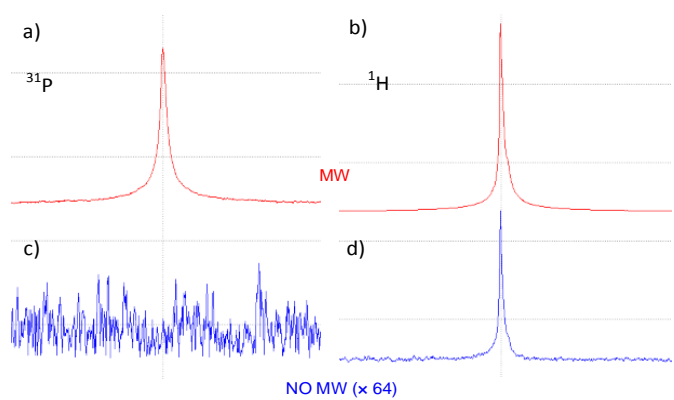

**Fig.4.** $^{31}$P and $^1$H spectra with 60 mM BDPA radical acquired simultaneously at one scan. (a) and (c) are $^{31}$P spectra with (red) and without(blue) MW irradiation. (b) and (d) are $^1$H spectra with (red) and without (blue) MW irradiation. The thermally polarized spectrum is multiplied by 64.

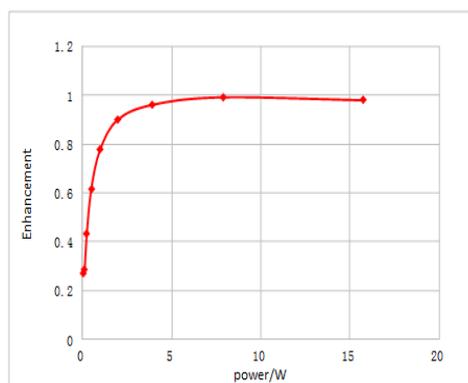

**Fig.5.** Normalized DNP enhancement factor ε of $^{31}$P as a function of the microwave power

## 3.2  $^1$H and $^{23}$Na spectra

Sodium ion homeostasis occurs between intracellular and extracellular sodium ions [25]. The intracellular sodium ion concentration is approximately 10 mM, much lower than the extracellular concentration of about 140 mM. In the case of diseases, these sodium ion concentrations are known to change and hence measurements of sodium concentration in tissue can provide information on the status of the tissue, potentially aiding in disease diagnosis or therapy monitoring. Fig.6 shows the simultaneously acquired $^1$H and $^{23}$Na signal of with DNP enhancement (Fig.6a and b) and without enhancement (Fig.6c and d). The time of microwave excitation is 1s and microwave power is about 30W (MW resonator input terminal). A very large signal enhancement of up to 80 is obtained for $^1$H. Unfortunately, we can not directly determine the DNP enhancement factor of $^{23}$Na since $^{23}$Na signal is undetectable without MW irradiation. Our next goal is to obtain the spectra or image of $^{23}$Na and $^{39}$K simultaneously, which is significant for clinical diagnostics and biomedical research.

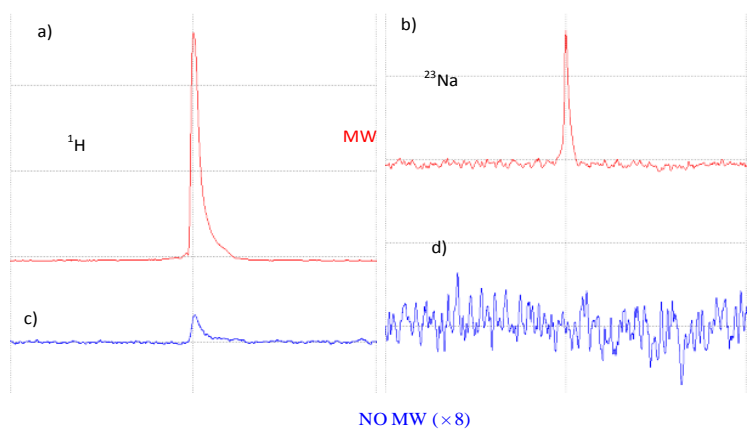

**Fig.6.** Simultaneously acquired $^1$H and $^{23}$Na spectra after one scan. (a) and (c) are $^1$H spectra with (red) and without (blue) MW irradiation. (b) and (d) are $^{23}$Na spectra with (red) and without (blue) MW irradiation. The thermally polarized spectrum is multiplied 8 times.

### 3.3 $^{19}$F and $^{31}$P spectra

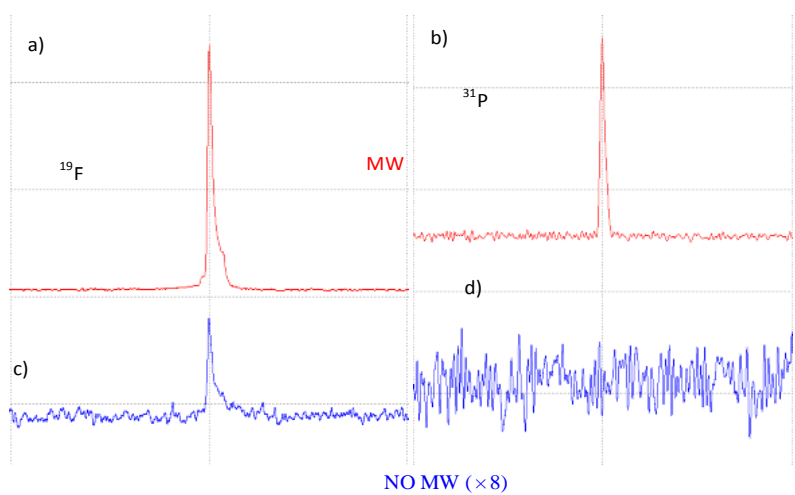

**Fig.7.** Simultaneously acquired $^{19}$F and $^{31}$P spectra with and without Overhauser enhancement for benzene/hexafluorobenzene/triphenyl phosphorous solutions. (a) and (c) are $^{19}$F spectra with (red) and without(blue) MW irradiation. (b) and (d) are $^{31}$P spectra with (red) and without (blue) MW irradiation. The thermally polarized spectrum is magnified 8 times.

Due to the relatively high sensitivity, no interfering background signals and much broader chemical shift range, $^{19}$F NMR spectroscopy has a potential application in molecular biology and biochemistry investigations including the structure and function of proteins and nucleic acids, enzymatic mechanisms, metabolic pathways and biomolecular interactions [26]. DNP enhanced $^{19}$F NMR could compensate the loss of sensitivity at lower fields as compared to high field measurements. Fig.7 shows the simultaneously acquired $^{19}$F and $^{31}$P spectra with enhancement (Fig.7a and b) and without enhancement (Fig.7c and d) at 0.35T. The microwave pumping time is 2s and microwave power is about 30W (resonator input terminal). Very strong $^{19}$F and $^{31}$P NMR signals are seen in the DNP-enhanced co-acquisition spectra, but in the co-acquisition spectra without DNP enhancement $^{19}$F signal is very weak and $^{31}$P signal is totally unobservable. $^{19}$F DNP enhancement is about 20.

### 3.4 $^1$H and $^{31}$P image

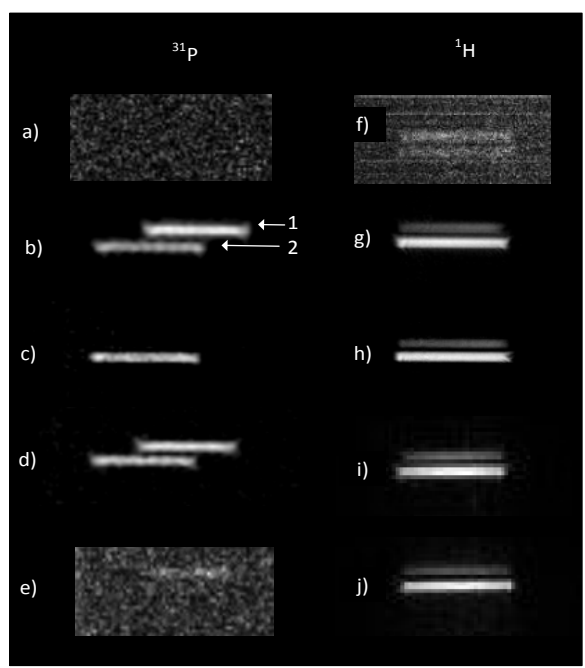

**Fig.8.** $^{31}$P (a-e) and $^1$H (f-j) images obtained at 0.35T. Sample 1: PCl$_3$ and benzene, sample 2: triphenyl phosphorous and benzene. a), b), c), and f), g), h) are acquired separately from $^{31}$P and $^1$H; d) and i) , e) and j) are from parallel acquisitions. Independent control of TE to obtain the image of T$_2$-weight, a), b), d), and f), g), i): TE=11ms; c), e), h), j): TE=35ms. Other experiment parameters: TR=3s, phase encode = 128, MW power = 8W.

An in-house-built user interface software, which supports experimental operations, data processing and graphic editing pulse sequences, was used for MRI. Because the same gradient and sampling bandwidth were utilized for spatial encoding in the $^{31}$P and $^1$H co-image (Fig.2b), the different nucleus spatial frequency sampling leads to different FOV and resolution. This difference in FOV and resolution can be corrected by scaling the k-space trajectory or interpolating data in image space during image reconstruction. Fig.8 shows a series of $^{31}$P and $^1$H images with (Fig.8b-e, g-j) and without (Fig.8a, f) microwave irradiation, in the cases of parallel acquisition (Fig.8d-e, i-j) and single-channel acquisition (Fig.8b-c, g-h). The numbers 1 and 2 in Fig.8 denote sample 1 and sample 2 respectively. One sees clearly the shape of the sample from DNP-enhanced images (Fig.8b and g), but can not discern the profile of the sample without DNP enhancement from either $^{31}$P or $^{1}$H (Fig.8a and f). Since the $^{31}$P $T_2$ of PCl$_3$ is much shorter than that of the triphenyl phosphorous ((C$_6$H$_5$)$_3$P, PPh$_3$), we can only observe the image of $^{31}$P in PPh$_3$ but not in PCl$_3$ (Fig.8c) when the echo time TE is long. Owing to the J-coupling effect between $^{1}$H and $^{31}$P in PPh$_3$, we can see the parallel images of $^{31}$P and $^{1}$H only in the case of short TE during which $^{31}$P evolution caused by $^{31}$P-$^{1}$H J-coupling in PPh$_3$ can be neglected. Whereas for long TE period during which a pair of π pulses of $^{1}$H and $^{31}$P was applied simultaneously and the J-coupling transforms the $^{31}$P in-phase coherence into the anti-phase coherence, the $^{31}$P image from parallel acquisition becomes unobservable in PPh$_3$ (Fig.8e). But the $^{1}$H image is still clear in the latter case because most of the protons are not J-coupled with the $^{31}$P. In addition, we can also obtain the chemical shift of $^{31}$P nuclei with different position shifting from the center of the image (Fig.8b).

### 3.5 $^{19}$F and $^{31}$P image

We have also simultaneously obtained the co-images of $^{19}$F and $^{31}$P of sample 4. Fig.9 shows the images of $^{19}$F and $^{31}$P with (Fig.9a, b) and without (Fig.9c, d) microwave irradiation. We see clearly the $^{31}$P and $^{19}$F images of the sample in the case of DNP enhancement (Fig.9a, b) but can not discern the profile of the sample without DNP enhancement from either $^{31}$P or $^{19}$F (Fig.9c, d).

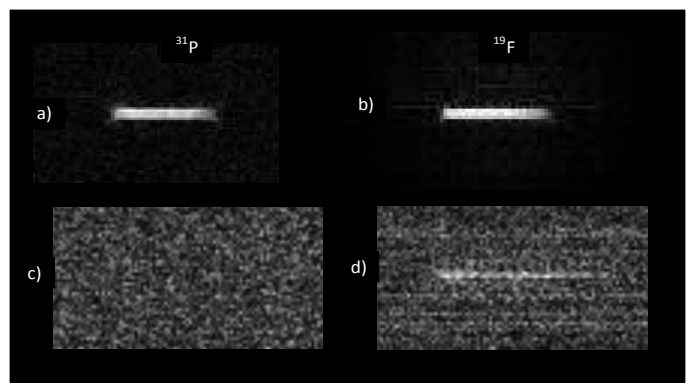

**Fig.9.** $^{19}$F (a,c) and $^{31}$P (b,d) co-images with BDPA radical from parallel acquisitions. Other experiment parameters: TR=3s, TE=40ms, phase encode = 128, MW power = 30W.

### 4. Discussion and Conclusion

In summary, we have realized the parallel spectra and images of two nuclei based on simultaneous multiple-nuclei Overhauser DNP enhancement. These DNP enhanced multi-nuclei NMR spectra/images may provide more significant information of metabolism *in vivo*. For example, we may simultaneously monitor the sodium ion transport and phosphorus metabolism by analyzing the spectral data. Without doubt, simultaneous acquisition of multi-nuclear enhanced magnetic resonance through DNP offers a number of advantages over classical simultaneous acquisition and DNP approaches. It could significantly broaden NMR applications in biomedical systems especially for studying low-γ nuclei which are important for metabolism. Structural information, ion transport pathway and metabolic processes can be simultaneously observed with various nuclear spectroscopy and image. The spin–spin coupled relationship between the different nuclei can be extracted either from two-dimensional and multidimensional spectra or image and used for three-dimensional structure refinement [14]. At the same time, it significantly reduces the acquisition time for multidimensional and quantitative analysis experiments [27]. This method can also be used for many fast acquisition techniques [28], for instance, Hadamard NMR spectroscopy [29], computer optimized aliasing [30] and projection reconstruction method of protein [31] with enhanced signal at low field. What's more, we can image with different nuclei at a single molecule of different position or at different molecules to obtain considerably complementary information. We can obtain two different images if the distribution of the sample is inhomogeneous or the chemical shift of nuclei can be resolved distinctly. This method can also be used in high field dynamic nuclear polarization to obtain the high-resolution and high-sensitivity of NMR spectra and image for biological samples [32-34].

## ACKNOWLEDGMENT

This work was supported by the Chinese Academy of Sciences (grant no. ZDYZ2010-2), the Ministry of Science and Technology of China (grant no. 2011YQ120035), and the National Natural Science Foundation of China (grant no. 11405264, 11274347, 21221064, 11575287).